\documentclass[letterpaper, 10 pt, conference]{ieeeconf}  

\IEEEoverridecommandlockouts                              

\usepackage{enumerate}
\usepackage{color}
\usepackage{graphicx}
\usepackage{cite}

\title{\LARGE \bf
Methodological Approach for the Design\\of a Complex Inclusive Human-Machine System
}

\author{Lorenzo Sabattini$^{1}$, Valeria Villani$^{1}$, Julia N. Czerniak$^{2}$, Alexander Mertens$^{2}$ and Cesare Fantuzzi$^{1}$
\thanks{$^{1}$L. Sabattini, V. Villani and C. Fantuzzi are with the Department of Sciences and Methods for Engineering (DISMI), University of Modena and Reggio Emilia, Reggio Emilia, Italy
        {\tt\small \{lorenzo.sabattini, valeria.villani, cesare.fantuzzi\}@unimore.it}}%
\thanks{$^{2}$J. N. Czerniak and A. Mertens are with the Institute of Industrial Engineering and Ergonomics, RWTH Aachen University, Aachen, Germany
        {\tt\small \{j.czerniak, a.mertens\}@iaw.rwth-aachen.de}}
}

\begin{document}

\maketitle
\thispagestyle{empty}
\pagestyle{empty}

\begin{abstract}

Modern industrial automatic machines and robotic cells are equipped with highly complex human-machine interfaces (HMIs) that often prevent human operators from an effective use of the automatic systems. In particular, this applies to vulnerable users, such as those with low experience or education level, the elderly and the disabled. To tackle this issue, it becomes necessary to design user-oriented HMIs, which adapt to the capabilities and skills of users, thus compensating their limitations and taking full advantage of their knowledge. In this paper, we propose a methodological approach to the design of complex adaptive human-machine systems that might be inclusive of all users, in particular the vulnerable ones. The proposed approach takes into account both the technical requirements and the requirements for ethical, legal and social implications (ELSI) for the design of automatic systems. The technical requirements derive from a thorough analysis of three use cases taken from the European project INCLUSIVE. To achieve the ELSI requirements, the MEESTAR approach is combined with the specific legal issues for occupational systems and requirements of the target users.

\end{abstract}

\section{Introduction}

Advances in technology in modern industrial settings have led to the introduction of extremely complex automatic machines and robotic cells.
Despite such a massive introduction of advanced technological solution, the role of human operators in this context is still focal, since they are responsible for controlling and supervising manufacturing activities and the desired flexible production. Nevertheless, this new technological scenario is not favorable to human operators themselves: indeed, the complexity of modern manufacturing plants is reflected in an increased complexity of the accompanying human-machine interfaces (HMIs), which allow the user to operate the machine, observe the system status and, if necessary, intervene in the process \cite{Skripcak_2013, Nachreiner_2006}. The increase in complexity of modern industrial HMIs can still be tackled by the most experienced human operators, who can interact efficiently with the machine only at the expenses of an unsustainably increased mental workload and stress. However, in the worst condition, vulnerable workers, such as those with low experience or education level, the elderly and the disabled, can barely sustain such an interaction in an effective manner.

To tackle this issue, it is needed to make use of an anthropocentric approach that reverses the paradigm from the current belief that "the human learns how the machine works" to the future scenario in which "the machine adapts to the human capability" accommodating to her/his own time and features \cite{Norman_2013}. This is realized by adaptively simplifying the HMI based on the user's features and complementing her/his cognitive capabilities by advanced sensing and higher precision of machines. Following such approach, it would be possible to create an \emph{inclusive} \cite{Abascal_2005, Stephanidis_2001} and flexible working environment for any kind of operator, taking into account multiple cultural backgrounds, skills, age and different abilities. 
 developing a methodology for the design of adaptive human-centered HMIs for industrial machines and robots.



HMIs typically used for supervising industrial processes do not provide any possibility of controlling the amount of displayed information, or its form. Hence, while the human operator is flexible and adaptable, the system is not. In particular, the control systems applied to industrial processes typically respond in a specified way, without regard as to whether the flow of information is low or extremely high, or the level of expertise of the user is good or bad~\cite{Viano_2000_short}. 
The human operator is then typically the only element that needs to adapt her/his behavior based on the situation. Namely, the operator needs to be sufficiently flexible, to be able to cope both with common activities and unpredictable situations, such as in the presence of dangers. This can cause significant difficulties for the operators, in particular considering the fact that the amount of monitored data that come from modern production processes is constantly increasing, and control systems are becoming increasingly complex \cite{Flaspoler_2010, Viano_2000_short, Skripcak_2013}. 

To overcome this issue, the concept of context-dependent automation, also known as adaptive automation, has been introduced~\cite{Parasuraman_2000, Lee_2013}. Generally speaking, context awareness is the ability for a system to sense, interpret, respond and act based on the context~\cite{Dey_2001}. Based on this paradigm, the level of automation of a system is designed to be variable, depending on situational demands during operational use.

Along similar lines, the idea of adaptive user interfaces has been developed, which consist in changing how the information is presented, in such a way that only the relevant pieces of information are provided to the operator, based on the context. Examples of adaptive user interfaces have been developed considering different application domains, such as automotive \cite{Sharma_2008, Garzon_2012}, aeronautics \cite{Inagaki_2000} and smartphones and hand-held devices \cite{Gu_2004}. However, to the best of the authors' knowledge, only a few pioneering examples have been preliminary presented regarding HMIs for complex industrial systems~\cite{Viano_2000_short, Lee_2013}. Specifically,~\cite{Viano_2000_short} described a preliminary concept of architecture for an HMI that adapts the presentation of information based on the operator responsiveness. Profiling of the operators is considered in~\cite{Lee_2013}, and the HMI selectively presents information based on the profile of the current user.

Going beyond this state of the art, the European project INCLUSIVE aims at developing a smart interaction system that adapts the information load of the HMI and the automation capability of the machine to the physical, sensorial and cognitive capabilities of workers \cite{Villani_2017_ETFA}. In particular, the final goal is to provide technological solutions for compensating workers' limitations (e.g. due to age or inexperience), while taking full advantage of their knowledge. Three groups of operators are considered, namely elder, disabled, and inexperienced operators, since they are believed to be the most vulnerable ones in the interaction with complex automatic systems, as discussed in Sec.~\ref{sec:use_cases}.

Three main pillars constitute the INCLUSIVE system \cite{Villani_2017_ETFA}.
The first pillar relates to the measurement of human capabilities: the system will measure the human capability of understanding the logical organization of information and the cognitive burden the operator can sustain (automatic human profiling).
The second pillar consists in the adaptation of interfaces to human capabilities: the system will adapt the organization of the information, the means of interaction, and the automation task that are accessible by the user depending on her/his measured capabilities.
Finally, the third pillar is about teaching and training for unskilled users: the system will be able to teach the correct way to interact with the machine to the unskilled users, exploiting also simulation in virtual and augmented environment.

In this paper, we propose a set of methodological recommendations for the design of an adaptive human-machine system that is inclusive for all users. In particular, we derive the technical requirements that a complex human-machine system, such as the one considered in INCLUSIVE, should fulfill in order to allow also vulnerable users to access it. Such requirements are defined starting from the analysis of the industrial use cases of INCLUSIVE, but have general validity. In particular, the main issues related to state of the art solutions in terms of HMI are highlighted, referring explicitly to representative target scenarios.
From the analysis of the use cases, a set of users' needs is defined. Specifically, users' needs describe the technical issues and difficulties that operators typically encounter with the currently available technological solutions.
Users' needs are then abstracted, to define the technical system requirements. These are general technical methodological guidelines that should be considered in the design of any complex human-machine system, in order to make it accessible also to vulnerable users. 
Moreover, we carry out an analysis of the different ethical, social and legal implications (ELSI) of such a system to protect the user against harm and disadvantages. Based on the MEESTAR approach \cite{Manzeschke_2015}, which is an instrument for identifying ethical problems, we develop an ELSI concept and test its appropriateness in a possible operative scenario. Then, we derive some design recommendations in terms of ELSI requirements for the development of smart interaction systems for automated production machines. The aim is offering fair requirements, independent of individual skills and capabilities.
%

\section{Description of the considered use cases}\label{sec:use_cases}
To derive methodological considerations that have general validity it is important to start from real use cases that depict the scenario of human-machine systems currently utilized in industrial environments. To this end, we consider, as a case study, the industrial use cases addressed in the INCLUSIVE project, since they are representative of a wide area of interest for industry in Europe:

\begin{enumerate}[\textit{Use case} \itshape1\upshape:]
	\item machinery for small companies, typically run by elderly owners;
	\item automation solutions made for developing countries;
	\item industrial plants made by a big company.
\end{enumerate}

Specifically, the first use case refers to machinery used for woodworking in artisans' shops. The second one considers a robotic solution to be applied in a company located in a developing country, where operations are mostly performed manually. In particular, the considered robotic solution is for panel bending. Finally, the third use case refers to a bottling company and, in particular, a labelling unit is considered.

Such use cases have been chosen since they address different categories of most vulnerable users, namely elderly, disabled and low experienced.
Specifically, by elderly we consider those people in the last years of their work life. Generally, these workers have a large experience in the traditional industrial processes, but are not familiar with modern computerized devices and, then, have difficulties in utilizing modern automatic machines that come with complex HMIs.
As regards people with physical impairment and limited cognitive abilities, such limitations introduce as well difficulties in the use of complex automatic machines.
Finally, by inexperienced we refer to people with low level of education, limited expertise in the use of automatic machines and/or computerized HMI, and lack of experience in industrial processes.

For each use case, a specific working scenario is analyzed in order to derive what are the concrete limitations of currently implemented solutions.
These activities were selected by the corresponding industrial partners of the INCLUSIVE consortium, since they require unavoidable interaction of the user with the machine and are representative of the most frequent operations with automatic machines.
Specifically, for the first use case we focus on the activities related to tuning of the machine, to make it ready for woodworking (tuning of the tools warehouse, tuning of the worktable area components) and routine maintenance procedures.
For the second use case, we consider the standard activities performed by a user for bending a part, and replacing malfunctioning tools. The working scenario for the third use case refers to the fault recovery procedure, performed in jog mode, for misalignment of the neck ring label of bottles and the changeover of the printing format, required at the beginning of a working day or when a new bottle or label is produced on the line.

\section{Analysis of the problems of current HMIs}\label{sec:main_issues}
For each of the working scenarios, we analyzed how interaction is currently carried out, aiming at finding pitfalls which should be corrected in an inclusive system. 

\subsection{Use case 1}
The first limitation in the current implementation of the human-machine interaction lies in the fact that there is a clear lack of guided procedures assisting the user. In fact, the user is currently barely supported by the interface: only simple alarms are displayed, which describe what the current problem is, but not how to solve it.
Moreover, as regards the setup of the tools change, there is a misalignment between the equipment in the physical store (i.e., the tools on board the machine) and that in the virtual one shown in the HMI (i.e., the tools that the HMI displays as on board the machine), since the virtual store does not update automatically when a change in the physical one is made. As a consequence, currently the operator must pay attention to avoid mistakes that could jeopardize the operation of the machine: clearly, this activity is time-consuming and prone to errors.

This consideration applies also to the setup of the working area. Indeed, currently the interface supports the operator only by displaying, in a picture, the position of the components. It is up to the operator to manually move the various components in the correct position.
This lack of intuitiveness and assistance results in an additional decrease of efficiency and raises problems related to the constant need to consult the operator's manual, thus stopping normal operations to solve routine issues. However, since the manual is typically not stored close to the machine and is not organized with a clear focus on troubleshooting, it is rarely used by the operators, who end up to directly contact the assistance service to solve routine issues. In some other cases, they perform some tasks following some unofficial shortcuts rather than the official procedures recommended in the manual.
Moreover, given the lack of guidance, often errors of inexperienced operators severely compromise the operation of the machine.

\subsection{Use case 2}
With respect to the second use case, the main problem of the current HMI lies in the fact that these robotic cells can be used only by highly skilled personnel. In particular, background education in mechanical or electrical fields is necessary, since operators need to have significant coding skills both to program the system, and to be able to recover from problems that could arise during normal operations, also for simple cases, such as photocell malfunction. The use of the system by unskilled operators usually causes several problems since they often choose the wrong tool to perform the bending operations, or the wrong material thickness, thus making bending not possible, or wrong settings in the definition of the air pressure, that thus leads to incorrect bending operations. Moreover, current HMIs are based on touch screens and standard computers, and they cannot be utilized by people with disabilities of the upper limbs, or by blind people, effectively. Further, as in the previous use case, no guided procedure is available, besides the manual: hence, only operators with a long experience are able to solve problems. Although several choices need to be made for setting up the system (e.g. the correct angle to be used for bending a certain part), commonly adopted solutions exist, but they are known only by expert operators. Also, the operator needs to decide what parameters need to be changed, and then see what the result will be: again, this operation is mainly based on the operator's experience.

\subsection{Use case 3}
As regards the last use case, one of the biggest issues is related, also in this case, to the fact that the use of the system by untrained users is impractical. Operators need a specific training phase, before being able to interact with the machine. In particular, during the first uses, operators perceive the interaction with the system as uncomfortable. In these conditions, it was reported that operators feel afraid of damaging the system, the machine or the product, especially if a trainer or a supervisor is close by: indeed, although these people are trying to help or prevent disasters, the employee is stressed by this situation even more. Moreover, another source of stress is the fact that operators do not receive any feedback or acknowledgement of performed activities, to help them to understand if they are doing well.

Also in this case, inexperienced operators need the manual to check for every possible fault cause and how to correct them. Despite of this, it still happens that often wrong operations are performed, or operations are not correctly performed according to the manual, and, in particular, often the wrong operational mode is selected, e.g. semiautomatic or manual instead of jog mode. All these issues appear, in particular, for operators that are new to machinery or for low educated people.

\section{Definition of users' needs}\label{sec:users_needs}

The users' needs have been identified from the above analysis of the problems of current interaction systems. 

The first category of users' needs refers to the inclusion of all users in complex human-machine systems. The system should be effectively usable by inexperienced operators, by operators with different age, level of work experience, namely novice users and expert operators, and education, and those with physical impairment. Specifically, the presence of an easily accessible guidance, which might exploit augmented reality for step-by-step guided procedures, could be a substantial advantage for unskilled operators, in order to make problem solving tasks accessible also to them. In this regard, programming by code writing, which is currently required in the scenario of the second use case, should not be necessary. As regards physical impairments, different disabilities might be typical, depending on the application scenario: as an example, in the case of woodworking machines, missing fingers have been reported as a typical impairment. 

Thus, a second group of users' needs rises: the organization of information should be user-oriented. This implies that, on the one side, procedures should adapt to the operator's skills, thus being sufficiently clear for unskilled operators and not too long-winded for the skilled operators. On the other side, the system should guide the operator during ordinary operations, such as setup or maintenance. A teaching module could be implemented, to suggest unskilled operators common practice solutions. As a consequence, specific prior training and studying the manual should not be necessary. Despite of this, it should be possible to perform operations in the correct sequence, according to the manual, by means of proper suggestions suitably provided by the HMI. This should be possible also for tunable procedures, where the system should suggest the operator what parameters need to be changed, based on the desired result. A solution for unskilled operators could be to provide suggestions on what parameters need to be changed, knowing how they influence the achieved result.

These users' needs lead to the consideration that human factors must be prioritized. Indeed, the system should be perceived as comfortable for all the users and the stress level during the use of the system should be low. In order for this to be achieved, the intervention of supervisors for assisting the operators should be avoided and operators should feel confident when using the system alone.

As a consequence of such an anthropocentric approach, the operator's performance should be automatically enhanced, in the sense that the operators should be enabled to perform the correct actions and choices. The number of errors should be reduced, while the execution time should be improved. Specifically, the correct operational mode and the correct value for critical parameters should be automatically selected. Also, the choice of wrong options should be prevented and the HMI should depict the actual current equipment and state of the machine.

Finally, some advanced technological solutions should be implemented to allow a smoother interaction with the machine. Specifically, hands-free interaction, such as speech recognition and synthesis, should be possible to enable the operators to interact with the machine when wearing gloves or protection equipment. Additionally, portable interfaces, such as wearable devices and augmented reality, should be available, to guide the operators in the working area.

\section{Technical requirements}\label{sec:technical_req}
Based on the description of the use cases and of the identified user issues, the following system requirements are derived. They describe how an adaptive human-machine system should be implemented in order to be inclusive for all users, and in particular elderly, disabled and low experienced users:

\begin{enumerate}[\textbf{T-R}\bf 1]
	\item The interface adapts to the level of skills of the operator.
	\item The system can be used by low educated operators.
	\item The system can be used by physically and cognitively impaired operators.
	\item The system can be used by people with low computer skills.
	\item The system enforces the correct procedures.
	\item The operator feels satisfied from the interaction experience.
	\item Interaction with the system generates a low level of stress for the operators.
\end{enumerate}

\section{Ethical, social and legal aspects}\label{sec:ELSI}

\begin{figure}
	\centering
	\includegraphics[width=.7\columnwidth]{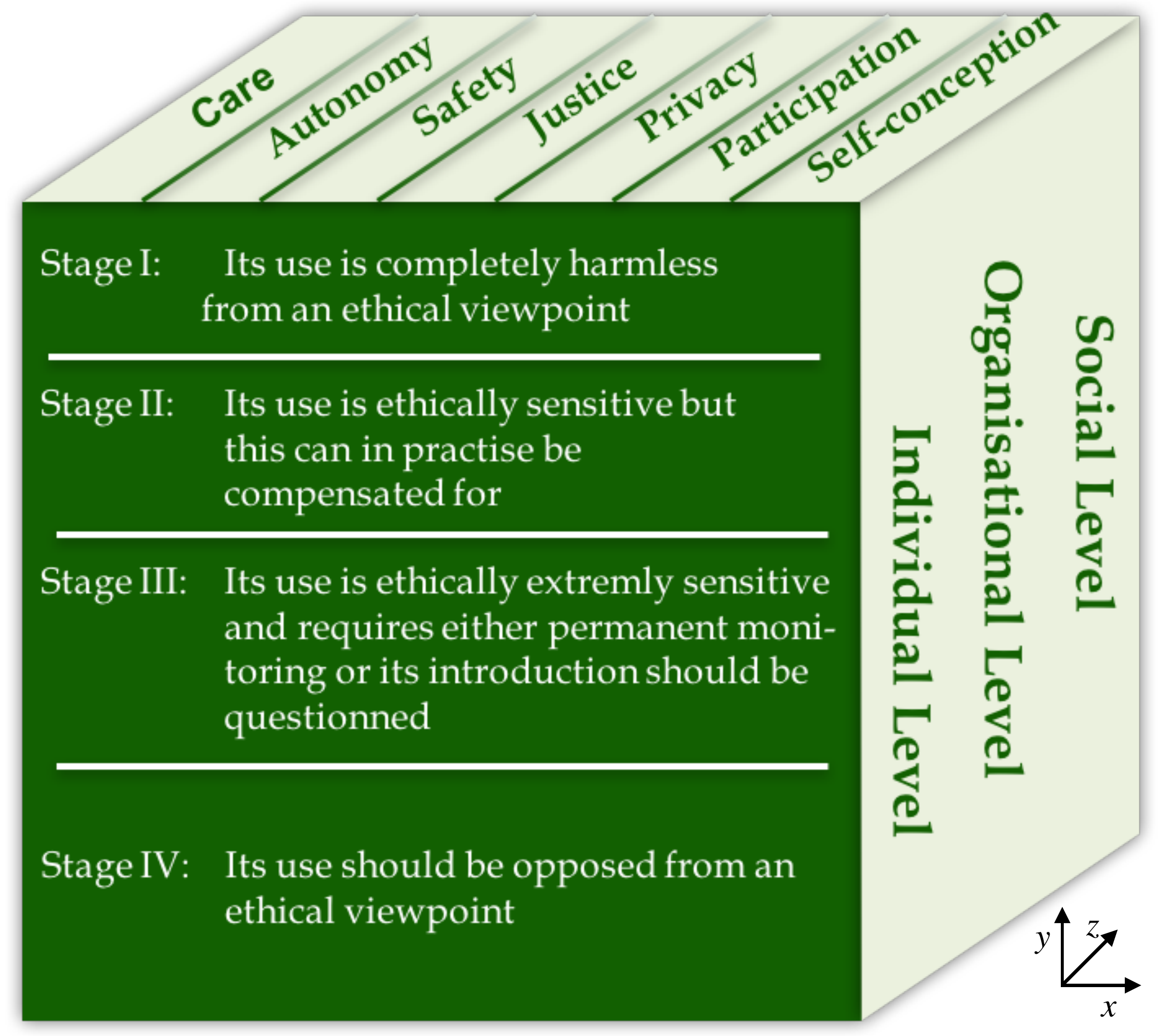}
	\caption{\label{fig:MEESTAR}MEESTAR model: $x$-axis: dimensions of ethical evaluation; $y$-axis: stages of ethical evaluation; $z$-axis: levels of ethical evaluation.}
	\vspace{-.2cm}
\end{figure}

The introduction of a system that processes sensitive personal data to disclose barriers of human capabilities, requires that also ethical, legal and social  requirements have to be taken into account to protect the user against harm and disadvantages. However, evaluating ethical, social and legal implications (ELSI) represents a specific challenge.
In this paper we propose to deal with ELSI aspects by a diverse approach, namely the MEESTAR model, which originally was developed for evaluating socio-technical arrangements in the field of age appropriate assisting systems \cite{Manzeschke_2015}. 
It is an analytical instrument which guides the process of reflecting on the use of technology. The model aims at identifying ethically problematic effects in a structured way and, on that basis, develop appropriate solutions. 
The model focuses on negative effects, requiring that the system causes little or no harm to the user. The first step of the MEESTAR analysis is to identify relevant ethical dimensions for the particular scenario. Thus, the aim of this approach is to find a basis for ethical, social and legal aspects, according to the intention of implementing sensors for measuring human capabilities and tracking individual health data. Furthermore, legal requirements given by the European Union (EU) are considered\footnote{In this paper we consider only EU legislation.}, and finally responsibility for needs of vulnerable target users is taken into account.

Working with MEESTAR involves the systematic consideration of three axes, as shown in Fig.~\ref{fig:MEESTAR}. The $x$-axis consists of seven ethical dimensions: care, autonomy, safety, justice, privacy, participation and self-conception. The $y$-axis describes stages of ethical evaluation, allocating problems among four levels of ethical sensitivity. The $z$-axis provides three points of view (individual, organizational, social).

The legal issues regard mainly data protection, safety and health at work, and product requirements. The main directives in the context of production machines are the Machinery Directive 2006/42/EC about construction of safety-related products, and the Council Directive 89/391/EEC on the introduction of measures to encourage improvements in the safety and health of workers at work.

\begin{figure}
	\centering
	\includegraphics[width=.95\columnwidth]{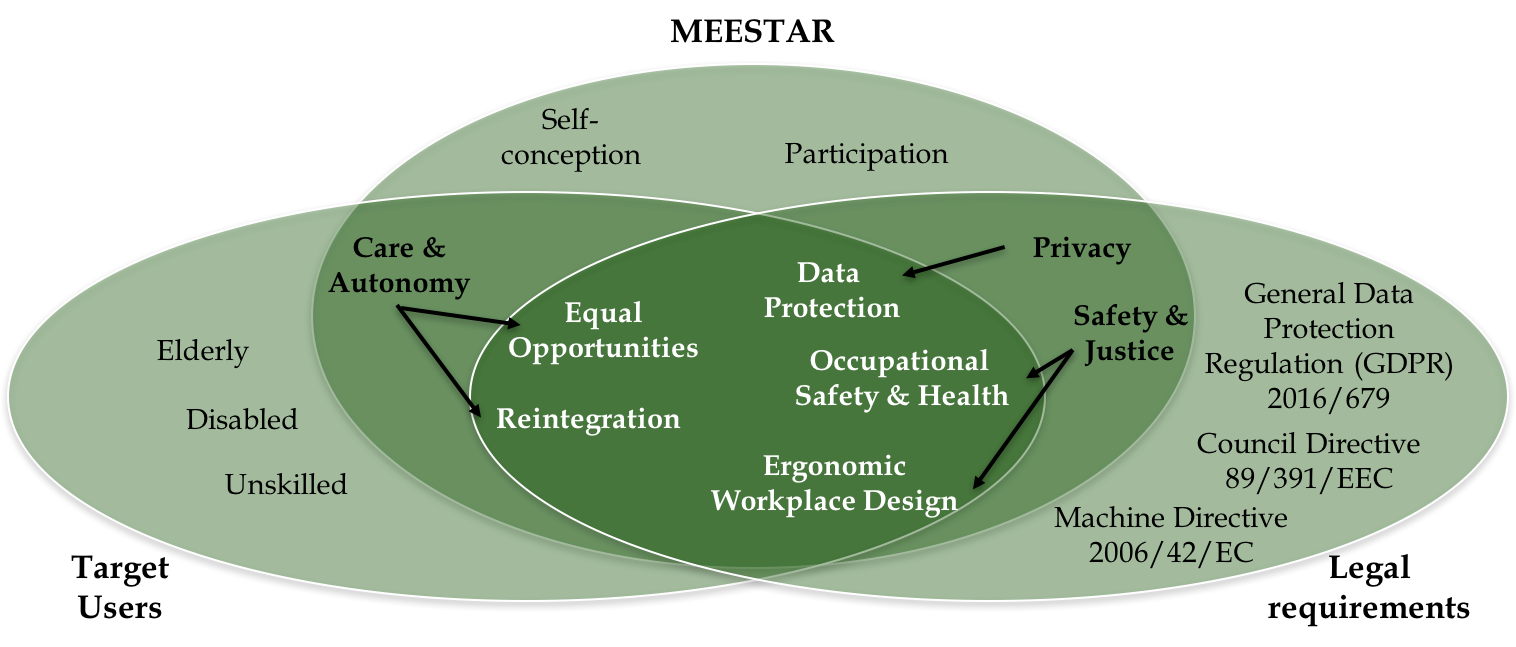}
	\caption{\label{fig:ELSI}The intersection of ethical, legal and social implications (ELSI) define the requirements for the considered inclusive human-machine system.}
\vspace{-.4cm}
\end{figure}

The MEESTAR dimensions show several intersections with legal requirements and target users, as shown in Fig.~\ref{fig:ELSI}:
\begin{itemize}
\item caring for users with different limitations in skills and capabilities,
\item giving these users possibility for an autonomous interaction with automated production systems,
\item fulfilling standards for safety and justice, by addressing employers corporate duties by law,
\item sensitively approaching the employees right to privacy according to legal requirements, by treating personal data with dignity and respect.
\end{itemize}

Thus, the following technical aspects that need to be taken into account can be derived: \textbf{i)} occupational health, \textbf{ii)} occupational safety, \textbf{iii)} data protection, \textbf{iv)} ergonomic workplace design, \textbf{v)} equal opportunities and \textbf{vi)} reintegration.
Specifically, occupational safety and health is an interdisciplinary field, concerning safety and health of a working person in an occupational system to prevent him/her from working hazards \cite{Hughes_2011} in accordance with the MEESTAR dimensions \textit{safety} and \textit{justice}. Also ergonomic workplace design, as a subtask of occupational safety and health promotion, belongs to this category. Under EU law, data can only be processed under strict conditions, because everybody has a right to the protection of personal data \cite{Weber_2010}, which corresponds to MEESTAR dimension \textit{privacy}. In the perspective of target users, who have special characteristics and therefore differences in perception, cognition and motor skills, an equal treatment and integration into working processes is required. Thus, \textit{care} about their capabilities and \textit{autonomous use} of automated machines are the main topics in this case.

\section{Assessment of the ELSI concept}
\begin{table}
	\centering
	\caption{\label{table:ELSI_potential_of_improvement}ELSI concept: potential of improvement and risks.}
	\includegraphics[width=.85\columnwidth]{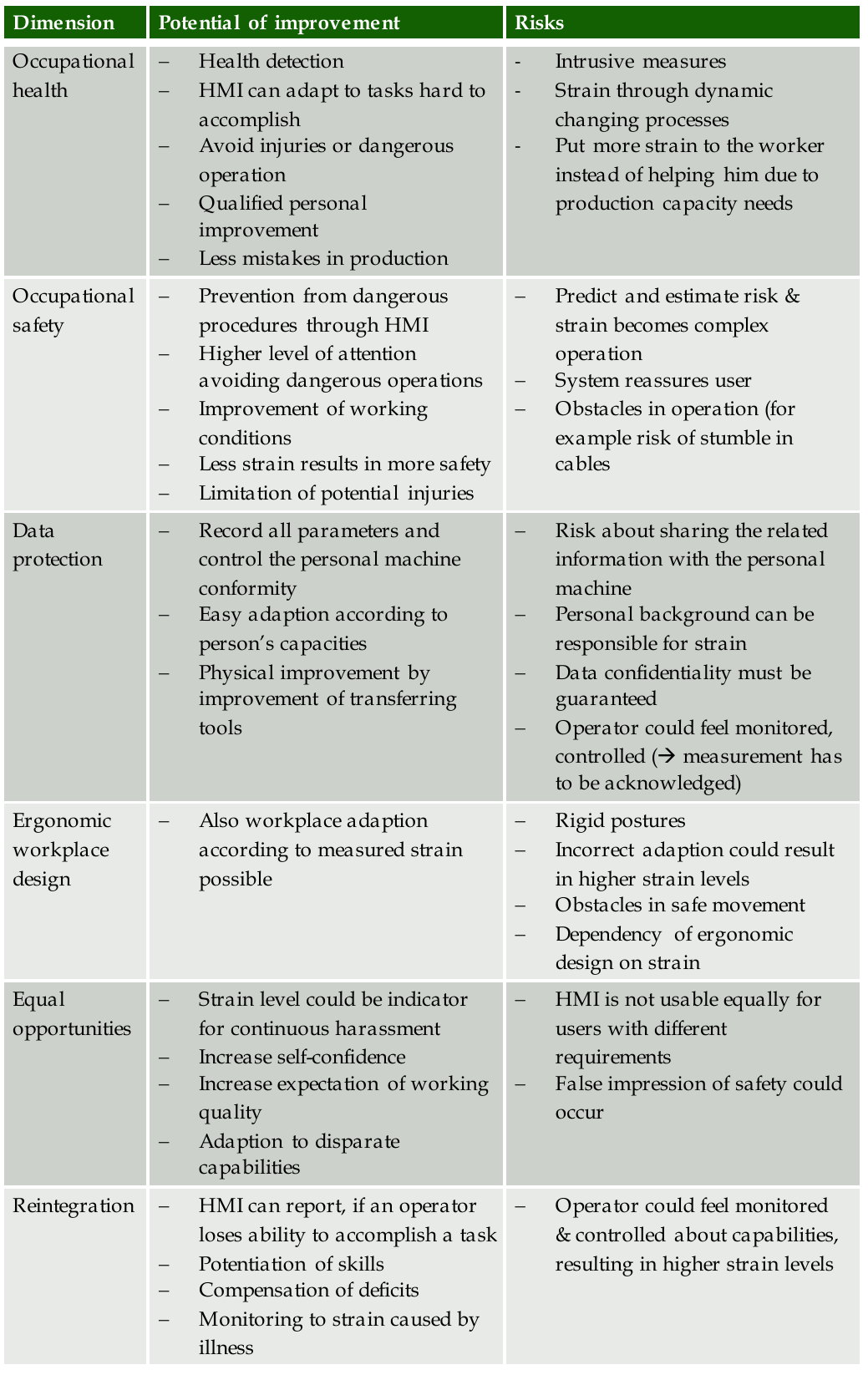}
	\vspace{-.3cm}
\end{table}
To assess the discussed dimensions of the ELSI concept, a questionnaire was developed to investigate the appropriateness of the identified dimensions in the considered scenario, namely that of an inclusive complex human-machine system accessible to special user groups, with special needs and requirements. 
To make the participation in the questionnaire more effective, we considered a specific working context where affective computing is applied to an industrial human-machine system, thus measuring operator's mental workload, stress and induced anxiety by recording some physiological signals. Specifically, the questionnaire included questions regarding the following scenario:
''\textit{The working machines are equipped with sensors that are able to track strain of a working person by real-time measurement of his/her physiological parameters, e.g. heart rate, blood pressure, etc. If the measured strain indicators are too high, the human-machine-interface adapts to the situation resulting in a lower stress level.}''

The questionnaire was distributed to all members of the INCLUSIVE consortium, to consider all relevant stakeholders that are affected. Seven partners participated in the study and participants were employed at companies in the following sectors: IT, technology transfer, industrial automation, white goods, packaging and bottling.
Each participant in the study was asked whether a potential of improvement/risks in measuring strain of a working person is measured according to each of the dimensions of the ELSI concept, namely occupational health, occupational safety, data protection, ergonomic workplace design, equal opportunities and reintegration.
Table~\ref{table:ELSI_potential_of_improvement} lists in detail all the potential of improvement and risks mentioned by the participants in the study. In particular, when designing the HMI, it has to be taken into account that the complexity resulting from the adaptive HMI behavior prevents inducing strain itself. In addition, the system must implement effective anonymization of personal user data; otherwise, there would be the risk that performance assessment, for instance, leads to a termination of employment. Moreover, the system should ensure that nobody is discriminated. According to respondents' answers, the supporting system should also ensure that the users have to respect safety regulations. Here, the system should meet relevant safety criteria and, if false impression of security occurs, call her/his attention. The measuring system should also take into account that the user is not distracted while working and that there is not a risk of stumbling. According to doubts of participants, the system should in no case cause injury to health by means of inductive measuring technology.


\subsection{ELSI requirements}\label{subsec:ELSI_req}

The findings reported in Table~\ref{table:ELSI_potential_of_improvement} allow us to derive the ELSI requirements, which have general validity and thus apply to any user-centred human-machine system that relies on affective computing for including vulnerable users.
Specifically, the derived design recommendations for ethical, social and legal aspects are the following:
\begin{enumerate}[\textbf{ELSI-R}\bf 1]
\item The system prevents inducing strain itself.
\item The system considers anonymized personal data.
\item The system uses collected data not for any disadvantage for the employer.
\item The system depicts relevant user requirements and prevents discrimination.
\item The system meets all relevant safety criteria.
\item The system does not distract the operator.
\item The system does not cause injuries by means of inductive measuring technology.
\end{enumerate}

\section{Conclusion}\label{sec:conclusion}
In this paper we presented a methodological approach to the design of complex human-machine systems that adapt to the operator’s skills and capabilities, complementing their limitations, while taking full advantage of their knowledge. Specifically, the proposed approach aims at guiding in the design of HMIs that can be effectively used by vulnerable operators, such as those with low experience or education level, the elderly and the disabled.
To this end, we defined a set of technical requirements and requirements related to ethical, legal and social implications. The technical requirements were derived from the analysis of the industrial use cases considered in the European project INCLUSIVE and they abstract what should be fulfilled in order to allow also vulnerable users to access a complex automatic machine or robotic cell were derived.
As regards the ethical, legal and social requirements, they were derived combining the MEESTAR approach 
with the specific legal issues for occupational systems and requirements of the target users. The validity of the such requirements was then validated in the context of the INCLUSIVE project.

\section*{Acknowledgement}
The research is carried out within the ''Smart and adaptive interfaces for INCLUSIVE work environment'' project, funded by the European Union's Horizon 2020 Research and Innovation Programme under grant agreement N°723373.
The authors would like to thank the industrial partners responsible for the use cases for providing a description of the use cases and selected working scenarios.

\bibliographystyle{IEEEtran}
\bibliography{INCLUSIVE}

\end{document}